\documentclass[pre,aps,twocolumn,superscriptaddress,showpacs]{revtex4}
\usepackage{amssymb,epsfig,amsmath,bm,subfigure,graphicx,textcomp,url,float,color,cases}
\usepackage{amsmath}

\begin{document}

\title{Near-field heat transfer between disordered conductors}

\author{Alex Kamenev}
\email{kamenev@physics.umn.edu}
\affiliation{Department of Physics and William I. Fine Theoretical Physics Institute, University of Minnesota, Minneapolis, MN 55455, USA}


\begin{abstract}
We study heat transfer mediated by near-field fluctuations of the electromagnetic field. In case of metals the latter are dominated by Coulomb interactions between thermal fluctuations of electronic density. We show that an elastic scattering of electrons, leading to diffusive propagation of density fluctuations,  results in a qualitative change of the radiation law. While the heat flux between clean metals follows the Stefan-Boltzmann-like $T^4$ dependence,  the heat exchange between disordered conductors is significantly enhanced and scales as $T^3$ at  low temperatures.

\end{abstract}

\maketitle

According to the Stephan-Boltzmann law the heat flux per unit area emitted by a black body with temperature $T$ is given by 
\begin{equation}
						\label{eq:Stephan-Boltzmann}
J_\mathrm{SB}(T)=\frac{c}{2} \int \! \frac{ d^3q}{(2\pi)^3} \, N(cq) =\frac{\pi^2}{60}\, \frac{T^4}{c^2}\,,
\end{equation}
where $N(\omega) =|\omega|/(\exp(|\omega|/T)-1)$ is the Planck function, $c$ is the speed of light and we work in the units where $\hbar=k_B=1$.   In particular, the heat flux between two closely spaced planar bodies kept at temperatures $T_1$ and $T_2$ is $J= J_\mathrm{SB}(T_1) -J_\mathrm{SB}(T_2)$, independent on distance $d$ between the two bodies. 

As was realized by Pendry \cite{Pendry1999} at small enough $d$ the heat flux may actually substantially exceed the Stephan-Boltzmann value (\ref{eq:Stephan-Boltzmann}) due to the presence of the near-field evanescent electromagnetic modes. Mahan has recently pointed out \cite{Mahan2017} that, if the two bodies are metals, such  near-field effect is dominated by the Coulomb interactions between thermal fluctuations of electron density. There is a number of recent experimental studies \cite{Reddy2015,Lipson2016}, which detect an excess heat flux between closely spaced parts of electronic nanostructures.    

The goal of this letter is to point out that the Coulomb-mediated heat transfer between the metals is highly sensitive to an  
elastic electron scattering. We show that at low temperatures, $T<1/\tau$, where $\tau$ is a mean scattering time, the heat flux scales as $T^3$, as opposed to Stephan-Boltzmann $T^4$ dependence (the latter scaling was found for near-field heat transport in clean materials, albeit with a $d$-dependent prefactor \cite{Pendry1999,Mahan2017,Kardar2017}).  Moreover, for a sufficiently large interlayer separation even at high temperature, $T>1/\tau$,  the heat flux is proportional to elastic scattering rate. We also find an intermediate temperature range, where the heat flux is universal and is independent of any material parameters.

The near-field heat flux between two metallic plates maintained at temperatures $T_1$ and $T_2$ is given by   
\begin{equation}
						\label{eq:heat-transfer}
J\!=\!\!\!
\int \!\! \frac{d\omega d^2q}{(2\pi)^3} 
[N_1\!(\omega)\!-\!N_{2\!}(\omega)] 
 \Im \Pi_1\!({ q},\omega)   \Im \Pi_2({ q},\omega) 
|U_{12}({ q},\omega)|^2\!\!,
\end{equation}
where $N_{i}(\omega)$ is the Planck function with temperature $T_{i}$, where $i=1,2$,  $\Pi_i({   q},\omega)$ 
is electronic density-density response function also known as polarization operator and $U_{12}({   q},\omega)$ is screened 
interlayer Coulomb interaction. In the random phase approximation the latter is found from the solution of the matrix Dyson equation:
\begin{equation}
				\label{eq:RPA}
\hat U({   q},\omega) = \hat U^{(0)}({   q}) +  \hat U^{(0)}({   q})\hat \Pi({   q},\omega) \hat U({   q},\omega), 
\end{equation}
where the polarization matrix is diagonal $\hat\Pi_{ij}=\Pi_i \delta_{ij}$ and for 2D planar geometry the bare Coulomb 
interactions are $\hat U_{11}^{(0)}=\hat U_{22}^{(0)}=2\pi e^2/q$ and  $\hat U_{12}^{(0)}=\hat U_{21}^{(0)}=2\pi e^2 e^{-qd}/q$. Here $d$ is the distance between the two conducting plates.  Solving Eq.~(\ref{eq:RPA}), one finds 
\begin{equation}
				\label{eq:U12}
U_{12}\! = \!\!\left[ 2\Pi_1\Pi_2\sinh(qd)\frac{2\pi e^2}{q}+\!\left(\Pi_1+\Pi_2+\frac{q}{2\pi e^2} \right) e^{qd} \right]^{-1}
\!\!\!\!\!.
\end{equation}
Equation~(\ref{eq:heat-transfer}) or its analogs have been derived for clean metals  \cite{Pendry1999,Mahan2017,Kardar2017,Wang2017,Wang2018}. 
Below we derive it for the diffusive case  employing kinetic equation approach \cite{Altshuler1979}. We notice the close resemblance 
of Eq.~(\ref{eq:heat-transfer}) to the expression for the Coulomb drag \cite{Solomon1990,Jauho1993,MacDonald1993,Kamenev1995}. The latter deals 
with the momentum transfer between the two layers, rather than the energy. However, unlike the drag conductivity, the energy flux (\ref{eq:heat-transfer}) is not limited to the linear response, i.e. it is not restricted to a small temperature difference.

To keep things simple we will restrict ourselves hereafter to two identical metals. If $T_{1,2}$ is smaller than the Fermi energy, the polarization $\Pi_{1,2}$ is approximately temperature independent and  Eq.~(\ref{eq:heat-transfer}) reduces to 
$$J=J(T_1)-J(T_2).$$ We notice, though, that such separation may not take place if the Fermi energy is small (as in  e.g. graphene \cite{Schutt2011}), leading to a $T$-dependent polarization.

\begin{figure}[htb]
\centering
  \begin{tabular}{@{}cccc@{}}
    \includegraphics[width=.45\textwidth]{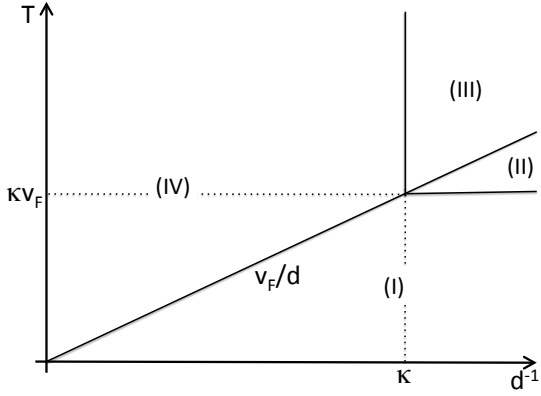}  
      \end{tabular}
  \caption{The four regions in temperature, $T$, vs. inverse  interlayer separation, $d^{-1}$, phase space. }
  \label{fig:Td-plane}
\end{figure}

We start from the clan case, $\tau^{-1}\to 0$, where the imaginary part of the polarization of a 2D metal is given by  \cite{Stern1967} 
\begin{equation}
					\label{eq:Pi-clean}
\Im \Pi({   q},\omega)=\nu\,\frac{\omega}{\sqrt{v_F^2q^2-\omega^2}}\, \theta(v_Fq-\omega), 					
\end{equation}
where $\nu$ is 2D density of states and $v_F$ is the Fermi velocity. For $v_Fq\geq \omega$ one has $\Re\Pi({   q},\omega)=\nu$. 
%
Depending on temperature and interlayer separation, one may identify four distinct regimes, which are marked in 
Fig.~\ref{fig:Td-plane}. In the low temperature regime (I) the dominant contribution to the integral is coming from 
$ \omega/v_F < q <\kappa/(1+\kappa d) \leq d^{-1}$ and $\omega\approx T$, where $\kappa=2\pi e^2 \nu$ is the inverse 2D screening radius. One may thus disregard the bare interaction term $q/(2\pi e^2)=\nu q/\kappa $ on the right hand side of Eq.~(\ref{eq:U12}), finding $U_{12}\approx   (2\nu)^{-1}/(1+\kappa d)$. Performing  2D momentum and energy integrations in Eq.~(\ref{eq:heat-transfer}), one finds:  
\begin{equation}
							\label{eq:J-clean}
J(T) =\begin{cases} 
\mbox{\large\(  \frac{\pi^2}{120} \frac{1}{(1+\kappa d)^2} \) \Large\(\,\frac{T^4}{v_F^2}\)\! \large\( \log\frac{\kappa v_F}{T(1+\kappa d)} \)} ,& \mathrm{(I)}\\
\\
\mbox{\large\( \frac{1}{12} \)} \kappa^2 T^2\,  \mbox{\large\( \log\frac{T}{\kappa v_F} \)}  ,  & \mathrm{(II)}\\
\\
\mbox{\large\( \frac{1}{8\pi^2} \, \frac{v_F}{d}\)} \kappa^2 T\, \mbox{\large\( \log\frac{1}{\kappa d} \)} ,  & \mathrm{(III)}\\
\\
\mbox{\large\( \frac{\pi^2}{1800} \frac{1}{(\kappa d)^2} \frac{v_F}{d^3}\)}\, T,   & \mathrm{(IV)}\\
\end{cases}
\end{equation}
In the high temperature regions (II) and (III), one may omit $\Pi_1\Pi_2$ term in Eq.~(\ref{eq:U12}). The heat flux is then primarily coming from  $\omega \approx T$ in (II),  and $\omega \approx v_F/d<T$ in (III).   Finally, in region (IV) $q\approx d^{-1}\ll \kappa$ and $\omega\approx v_F/d\ll T$, as a result $U_{12}$ is well approximated by the first term on the right hand side of Eq.~(\ref{eq:U12}), where one has to keep both real and imaginary parts of $\Pi_{1,2}$.  

Notice that, up to the logarithmic factor, the low temperature regime (I) mirrors the Stephan-Boltzmann $T^4$ dependence (\ref{eq:Stephan-Boltzmann}), though with the significantly larger coefficient for $d<\kappa^{-1}(c/v_F)$.  At larger temperature $T>v_F/d$, the $T^4\log(T)$ dependence gives way to the linear temperature dependence.
It is also worth noticing that in the regions (II) $J(T)\propto \kappa^2 T^2\log T$. This may be compared with the Pendry's upper bound   \cite{Pendry1999}: $J\leq   \pi T^2/3$ per ''information channel''. Though the strict definition of the latter is not clear, one may argue that the screening radius $\kappa^{-1}$  determines the minimum size of such a channel. If indeed, the regime (II) exceeds the Pendry's information bound. This fact is probably of little practical importance, since it only takes place when the distance between the two metals is less than the channel crossection. If distance $d$ is increased, one moves to regions (III) and (IV), where $J$ rapidly decreases below the bound. 

Let's briefly discuss  consequences of these findings for the interlayer temperature relaxation rate. For time dependence of temperature differential $\delta T = T_1-T_2\ll T_{1,2}$ one finds: $d\delta T/dt   = -\delta T/\tau_T$, with
\begin{equation}
						\label{eq:rate}
\frac{1}{\tau_T} = \frac{dJ(T)/dT}{C(T)}, 						
\end{equation}
where $C(T)=\pi^2\nu T/3$ is electronic specific heat per unit area. The low temperature $T^4\log T$ dependence of the heat flux, thus translates into $\tau_T^{-1}\propto T^2\log T$ relaxation rate. For $\kappa d <1$ this is nothing else but the celebrated  Giuliani and  Quinn \cite{Giuliani1982} 2D energy relaxation rate. Indeed, at small separation the interlayer relaxation rate must match the intralayer one. The interesting fact is that  the Giuliani-Quinn temperature scaling persists to large separations (though the rate is suppressed as $d^{-2}$) as long as $T<v_F/d$. At a higher temperature  $\tau_T^{-1}\propto T^{-1}$, because of the growth in the specific heat. This indicates that other mechanisms of the heat transport (e.g. phonons) eventually become more important at elevated temperatures.

\begin{figure}[t]
\centering
  \begin{tabular}{@{}cccc@{}}
    \includegraphics[width=.45\textwidth]{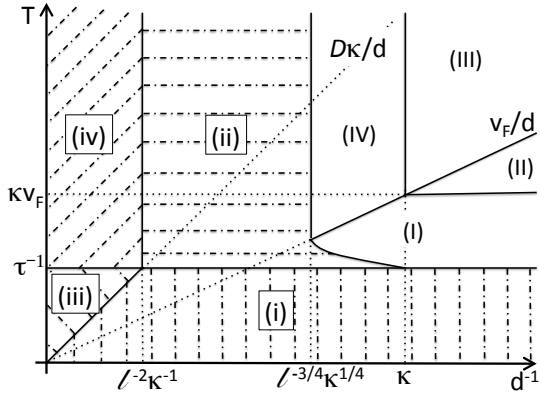}  
      \end{tabular}
  \caption{Temperature, $T$, vs. inverse  interlayer separation, $d^{-1}$, phase space in presence of disorder. The effects of disorder determine the heat flux in the shadowed part of the diagram ($l=v_F\tau$ is the mean free path). }
  \label{fig:Td-disorder}
\end{figure}

We turn now to our main subject -- disordered metals. Disorder changes surprisingly large part of the phase diagram, Fig~\ref{fig:Td-disorder}. As expected,  it's low temperature part, $T<1/\tau$, is qualitatively changed by the presence of disorder. The less evident feature is that also the high temperature part is affected at larger interlayer separations. The latter is due to extremely rapid decay, $\sim d^{-5}$, of the clean contribution in the high temperature 
regime (IV), Eq.~(\ref{eq:J-clean}).   Such a high power of the interlayer separation originates from the constraints due to 2D momentum conservation. Once the latter are relaxed by even a single scattering event at an impurity (taking place with the rate $1/\tau$), the heat flux decreases much slower, $\sim d^{-2}$.

Disorder leads to a diffusion propagation of the density fluctuations, resulting in the diffusion form of the polarization for small energy and momenta: $\omega<1/\tau$ and $q< 1/l=1/(v_F\tau)$  \cite{Altshuler1985,Kamenev2011} 
\begin{equation}
					\label{eq:Pi-diffusive}
\Pi({   q},\omega)=\nu\,\frac{Dq^2}{Dq^2-i\omega}, 					
\end{equation}
where $D=v_F^2\tau/2$ is the diffusion constant.  At not too large separation $d$ the interlayer interactions is dominated by the screening.  One may disregard the bare interaction on the 
r.h.s. of Eq.~(\ref{eq:U12}) and put $qd\ll 1$. This leads to:   
\begin{equation}
				\label{eq:U12-diffusive}
U_{12}({   q},\omega) = \frac{1}{2\nu} \frac{[Dq^2-i\omega]^2}{Dq^2[(1+\kappa d)Dq^2-i\omega]}.
\end{equation}
Substituting this in Eq.~(\ref{eq:heat-transfer}), one finds
\begin{eqnarray}
J(T)&=&\int\limits_0^{1/\tau}\!\!  \frac{d\omega}{\pi}\, N(\omega)\!\! \int\limits_0^\infty\! \frac{qdq}{8\pi}  \frac{\omega^2}{(1+\kappa d)^2(Dq^2)^2+\omega^2} \nonumber\\
&=& \frac{1}{(1+\kappa d) D} \int\limits_0^{1/\tau}\! \frac{d\omega}{8\pi}\,N(\omega)\,\omega  . 
								\label{eq:J-disorder}
\end{eqnarray}
Performing the frequency integration, one finds for the regions (i) and (ii) in Fig.~\ref{fig:Td-disorder}:
\begin{equation}
							\label{eq:J-disorder12}
J(T) =\begin{cases} 
\mbox{\large\(  \frac{\zeta(3)}{4\pi} \frac{1}{(1+\kappa d)D} \) } T^3 ,& \mathrm{(i)}\\
\\
\propto \mbox{\large\(  \frac{1}{(\kappa d)} \frac{1}{l^2\tau}\)}\,\, T,   & \mathrm{(ii)}\\
\end{cases}
\end{equation}
where $\zeta(3)\approx 1.202$ is Riemann zeta-function. Notice that the low temperature heat flux acquires a qualitatively  different scaling:  $T^3$, as opposed to $T^4\log T$ in the clean limit. This may be directly traced to the diffusive form of the polarization operator, Eq.~(\ref{eq:Pi-diffusive}). At higher temperature $T>1/\tau$ (regime (ii)) the heat flux is still dominated by the frequency $\omega\approx 1/\tau$. This is due to the fact the clean contribution (IV) in Eq.~(\ref{eq:J-clean}), coming from the higher frequency $\omega\approx v_F/d>1/\tau$, is too small, because of the rapid, $\sim d^{-5}$, decrease with the separation. 
%

At even larger separation $d>l^2\kappa$ the momentum integral in Eq.~(\ref{eq:J-disorder}) is limited by 
$q\leq1/d$ due to the exponential factor in the interlayer interaction potential (\ref{eq:U12}). This leads to additional  two regimes in Fig.~\ref{fig:Td-disorder}, where
\begin{equation}
							\label{eq:J-disorder34}
J(T) =\begin{cases} 
\mbox{\large\( \frac{1}{192}\,\frac{1}{d^2} \)}\,  T^2\,  ,  & \mathrm{(iii)}\\
\\
\propto \mbox{\large\(  \frac{1}{d^2\tau}\)}\, T.    & \mathrm{(iv)}
\end{cases}
\end{equation}
Remarkably in the intermediate regime (iii) bounded by $D\kappa/d < T<1/\tau$, the heat flux is completely {\em universal}, i.e. independent of {\em any} microscopic parameters of the material.  

The interlayer temperature relaxation rate, Eq.~(\ref{eq:rate}), at low temperature, (i),  is $\tau_T^{-1} \approx 0.087\, 
T/g(1+\kappa d)$, where $g=\nu D$ is dimensionless 2D conductance. For $\kappa d < 1$ this is nothing but Altshuler-Aronov \cite{Altshuler1979} relaxation rate of disordered metals. Similarly to the Giuliani-Quinn clean case, the Altshuler-Aronov low-temperature scaling persists to arbitrary separations, while the amplitude decreases as $d^{-1}$ (cf. with $d^{-2}$ in the clean case).  

The intermediate regime (iii) offers a peculiar example of a temperature-independent relaxation rate  
\begin{equation}
				\label{eq:bound}
\frac{1}{\tau_T}\propto \frac{1}{\nu d^2}, 				
\end{equation}
that is the mean level spacing in a square of size $d$. One my check that in all regimes marked in Figs.~\ref{fig:Td-plane}, \ref{fig:Td-disorder} the relaxation rate parametrically does not exceed  this value. Equation (\ref{eq:bound}) thus provides a universal (disorder independent) upper estimate for the interlayer temperature relaxation rate. This bound is somewhat similar in spirit (though not identical) to Pendry's information bound \cite{Pendry1999}.

In the remainder of the letter we outline derivation of Eq.~(\ref{eq:heat-transfer}) for the case of disordered metals. We start from 
the kinetic equation written for the distribution function $F(\epsilon; r,t)=1-2n(\epsilon; r,t)$, where $n(\epsilon; r,t)$ is the local occupation number \cite{Kamenev2011},  
\begin{equation}
						\label{eq-suppl:kinetic}
\partial_t 	F(\epsilon; t) = I^\mathrm{coll}[F].					
\end{equation}
We assume spatially homogeneous setup and thus disregard an $r$-dependence. Notice that, due to the fast elastic scattering, the distribution function depends only on energy $\epsilon$, rather than a momentum state $k$.  The heat flux is defined as time derivative of the total energy density
\begin{equation}
									\label{eq-suppl:flux}
J=\partial_t\left[ \!\!\int \!\! d\epsilon\, \nu(\epsilon) \,\epsilon\, n(\epsilon; t)\right] = -\frac{\nu}{2}  \int \!\!d\epsilon\, \epsilon\, I^\mathrm{coll}[F], 
\end{equation}
where we have assumed an approximately constant density of states at the Fermi energy $\nu(\epsilon)\approx \nu(\epsilon_F)=\nu$.

The collision integral for disordered metals was derived by Altshuler and Aronov  \cite{Altshuler1979,Altshuler1985,Kamenev2011} 
\begin{eqnarray}
&&I^\mathrm{coll}[F] =\! \frac{i}{2} \!\! \int \!\! \frac{d\omega d^2q}{(2\pi)^3}\,  \Re\, {\cal D}^R(q,\omega)
\big[(F(\epsilon-\omega)-F(\epsilon)) U^K \nonumber     \\ 
&&+ (1-F(\epsilon-\omega)F(\epsilon)) (U^R-U^A)    \big], 
						\label{eq-suppl:collision}
\end{eqnarray}
where $U^{R,A,K}=U^{R,A,K}_{11}(q,\omega)$ are retarded, advanced and Keldysh components of the {\em intra}layer interaction potential and ${\cal D}^R(q,\omega) = (Dq^2-i\omega)^{-1}$ is the diffusion propagator, originating from the renormalization of the interaction amplitude by the disorder.  We now take into account that $\hat U^{R,A} = 
( (\hat U^{(0)})^{-1} + \hat \Pi^{R,A})^{-1}$ to write $\hat U^R-\hat U^A = -\hat U^R(\hat \Pi^R-\hat \Pi^A)\hat U^A$. In the same way 
$\hat U^K = -\hat U^R\hat \Pi^K \hat U^A$. For the intralayer components one thus finds:
\begin{eqnarray}
&& U^R_{11}- U^A_{11} = - U^R_{11} (\Pi^R_1-\Pi^A_1) U^A_{11} - U^R_{12} (\Pi^R_2-\Pi^A_2) U^A_{21};  \nonumber \\
&&U^K_{11} = - U^R_{11} \Pi^K_1  U^A_{11} - U^R_{12} \Pi^K_2  U^A_{21}.
						\label{eq-suppl:RPA}
\end{eqnarray}
The first terms on the right hand sides here refer to the intralayer energy exchange. They lead to rapid thermalization and 
establishment of a local temperature $T_i$ within the layer $i$. Within this local equilibrium picture they do {\em not} contribute to the interlayer heat flux. We thus keep only the second terms on the right hand sides of 
Eqs.~(\ref{eq-suppl:RPA}). We also use that in local equilibrium $F_i(\epsilon) = \tanh \epsilon/2T_i$  thus $ (1-F_1(\epsilon-\omega)F_1(\epsilon))=(F_1(\epsilon)-F_1(\epsilon-\omega))\coth\omega/2T_1$, while 
$\Pi^K_2=(\Pi^R_2-\Pi^A_2)\coth\omega/2T_2$. As a result the interlayer part of the collision integral takes the form 
\begin{eqnarray}
&&I^\mathrm{coll}[F] =\! \frac{i}{2} \! \int \!\! \frac{d\omega d^2q}{(2\pi)^3}\,  \frac{Dq^2}{(Dq^2)^2+\omega^2} (\Pi^R_2-\Pi^A_2)|U^R_{12}|^2 \nonumber \\
&&\times\Big[\coth\frac{\omega}{2T_2} - \coth\frac{\omega}{2T_1}\Big] \big(F_1(\epsilon-\omega)-F_1(\epsilon)\big)  . 
						\label{eq-suppl:collision1}
\end{eqnarray}
We now substitute this into Eq.~(\ref{eq-suppl:flux}), perform the energy integration with the help $\int\! d\epsilon\, \epsilon 
 (F_1(\epsilon-\omega)-F_1(\epsilon)) = -\omega^2 $, employ $\omega (\coth\omega/2T_2-\coth\omega/2T_1)= 2(N_2(\omega)-N_1(\omega))$ and use Eq.~(\ref{eq:Pi-diffusive}) to write $\nu\omega Dq^2/(D^2q^4+\omega^2)=\Im\Pi_1^R$. This way we arrive at Eq.~(\ref{eq:heat-transfer}), where we suppressed superscript $R$ for brevity. 
 A similar procedure works to derive Eq.~(\ref{eq:heat-transfer}) also in the clean case. 
  
I am grateful to Igor Gornyi for useful discussions.  This work was supported by NSF grant DMR-1608238.

\end{document}